\documentclass[
aps,
prl,
reprint,
superscriptaddress,
amsmath,
amssymb,
]{revtex4-1}

\usepackage{color}
\usepackage{amsmath}
\usepackage{amssymb}
\usepackage{amsfonts}
\usepackage{amsthm}
\usepackage{float}
\usepackage{bm}
\usepackage{latexsym}
\usepackage{hyperref}
\usepackage{multirow}
\usepackage[capitalize]{cleveref}
\usepackage{graphicx}
\hypersetup{
    colorlinks,
    citecolor=blue,    filecolor=blue,
    linkcolor=blue,     urlcolor=blue
}

\begin{document}

\title{Ion acceleration with an ultra-intense two-frequency laser tweezer}

\author{Y. Wan}
\affiliation{Department of Engineering Physics, Tsinghua University, Beijing 100084, China}
\affiliation{Department of Physics of Complex Systems, Weizmann Institute of Science, Rehovot 7610001, Israel}

\author{I. A. Andriyash}
\affiliation{Department of Physics of Complex Systems, Weizmann Institute of Science, Rehovot 7610001, Israel}

\author{C. -H. Pai}
\affiliation{Department of Engineering Physics, Tsinghua University, Beijing 100084, China}

\author{J. F. Hua}
\affiliation{Department of Engineering Physics, Tsinghua University, Beijing 100084, China}

\author{C. J. Zhang}
\affiliation{University of California Los Angeles, Los Angeles, CA 90095, USA}

\author{F. Li}
\affiliation{University of California Los Angeles, Los Angeles, CA 90095, USA}

\author{Y. P. Wu}
\affiliation{Department of Engineering Physics, Tsinghua University, Beijing 100084, China}

\author{Z. Nie}
\affiliation{Department of Engineering Physics, Tsinghua University, Beijing 100084, China}

\author{W. B. Mori}
\affiliation{University of California Los Angeles, Los Angeles, CA 90095, USA}

\author{W. Lu}
\email[]{weilu@tsinghua.edu.cn}
\affiliation{Department of Engineering Physics, Tsinghua University, Beijing 100084, China}

\author{V. Malka}
\affiliation{Department of Physics of Complex Systems, Weizmann Institute of Science, Rehovot 7610001, Israel}
\affiliation{Laboratoire d'Optique Appliqu\'{e}e, Ecole polytechnique - ENSTA - CNRS - Institut Polytechnique de Paris, 828 Boulevard des Mar\'{e}chaux, 91762 Palaiseau Cedex, France}

\author{C. Joshi}
\affiliation{University of California Los Angeles, Los Angeles, CA 90095, USA}

\begin{abstract}
Ultra-intense lasers produce and manipulate plasmas, allowing to locally generate extremely high static and electromagnetic fields. This Letter presents a concept of an ultra-intense optical tweezer, where two counter-propagating circularly polarized intense lasers of different frequencies collide on a nano-foil. Interfering inside the foil, lasers produce a beat wave, which traps and moves plasma electrons as a thin sheet with an optically controlled velocity. The electron displacement creates a plasma micro-capacitor with an extremely strong electrostatic field, that efficiently generates narrow-energy-spread ion beams from the multi-species targets, e.g. protons from the hydrocarbon foils. The proposed ion accelerator concept is explored theoretically and demonstrated numerically with the multi-dimensional particle-in-cell simulations.
\end{abstract}

\maketitle
%\normalfont

While low-intensity laser tweezers and lattices manipulate microscopic objects \cite{ashkin1986,chu1991,grier2003}, lasers with higher intensities ionize matter and exert a force on the released plasma \cite{mourou2006optics}. In the subcritical density plasmas, where electron plasma frequency $\omega_{pe}= \sqrt{4\pi r_e c^2 n_e/\gamma_e}$ is smaller than that of the laser $\omega_0$, laser beams can propagate and drive wakefields, which is actively used now for electron acceleration \cite{RosenbluthPRL1972,tajima:PRL1979}. For the relativistically intense lasers, ${a_0=eE_L/m_ec\omega_0 \gg 1}$, electron plasma frequency is reduced by the Lorentz factor, $\gamma_e\approx a_0/\sqrt{2}$, allowing light to penetrate even solid, overcritical plasma densities \cite{akhiezer1956theory}. Such interaction occurs in the process of laser acceleration of ions, where laser heats a solid target to generate plasma with a strong charge separation \cite{daido2012review, macchi2013ion}. Here we have used $e$, $m_e$ and $r_e$ to denote the electron charge, rest mass and classical radius, $n_e$, and $\gamma_e$  stand for the density and effective Lorentz factor of electron plasma, and $c$ is the speed of light.

\begin{figure}[!ht]
\centering
 \includegraphics[width=\linewidth]{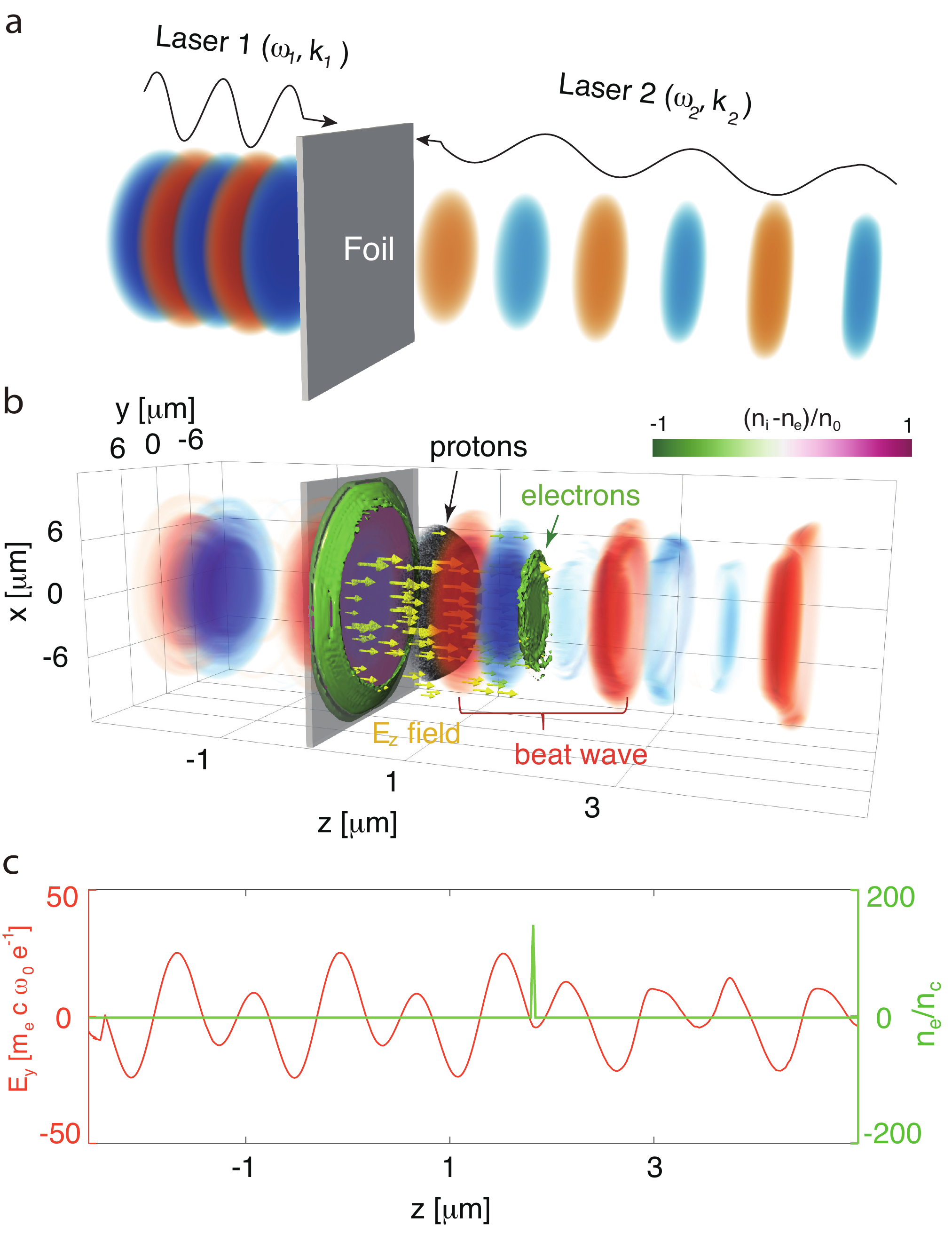}
 \caption{%\textbf{Three dimensional particle-in-cell simulation of relativistically intense two-frequency laser tweezer}.
 (a) Two circularly polarized lasers (electric field contours shown in red and blues) propagating towards a nano hydrocarbon foil (gray).
 (b) Interaction at $t=20$ fs: the extracted electron disk (negative charge shown with green) is followed by accelerated protons (black dots), while carbon ions are still immobile (positive charge is purple). Yellow arrows show a strong electrostatic field $E_z$ created between electrons and carbon ions.
 (c) Electron density $n_e$ and electric field $E_y$, on laser axis ($t=20$ fs).}\label{fig1}
\end{figure}

Modern high power lasers and advanced targetry provide a number of laser ion acceleration methods including acceleration via shealth fields \cite{snavely2000TNSA}, radiation pressure \cite{esirkepov2004highly, macchi2005}, collisionless shock \cite{silva2004, haberberger2012, fiuza2012prl}, magnetic vortices \cite{bulanov2005, bulanov2010pop, nakamura2010} etc. In most of the existing schemes, the charge-separation fields are produced by thermal electrons heated by intense lasers, that accelerate ions from solid targets, delivering an ion flux with a complex and typically broad spectrum of different species. The radiation pressure schemes, in particular the so-called light sail (LS) acceleration, take another path -- laser pushes an ultra-thin foil and accelerates all particles at the same rate, and thus is capable of generating quasi-monoenergetic beams \cite{esirkepov2004highly, klimo2008, robinson2008, yan2008, macchi2009}. In LS, the laser plasma heating is minimized. However, target becomes susceptible to the destructive plasma instability \cite{pegoraro2007, klimo2008}, which was shown recently to be related to the coupling between electron oscillations and ion plasma modes \cite{wan2016, wan2018physical}.  In this Letter we present a new acceleration concept, where two ultra-intense laser beams are used to spatially separate the electron and ion species thereby acting as a \textit{relativistically intense optical tweezer}. In this approach, plasma instability is fully mitigated, and high gradient acceleration is produced by the optically controlled charge-separation, which provides high quality of accelerated ions.

The proposed optical scheme relies on two circularly polarized (CP) lasers of different frequencies colliding on a single nano-foil (see Fig.\ref{fig1}). In order for the lasers to interfere inside the target, its thickness needs to be smaller than the relativistic skin depth of electron plasma, ${l_s=c\sqrt{\gamma_e}/\omega_{pe}}$. Once lasers collide on the target, their fields sum up into a beat wave, and its propagation direction and velocity are controlled by the ratio of the laser frequencies. Electrons of the foil first get squeezed into a thin sheet by the radiation pressure, and then are dragged by the beat wave, as if grabbed by a pair of tweezers. This creates a nearly constant longitudinal electric field between the electron sheet and the slow ion core (see Fig.~\ref{fig1}b and c). If the target contains an admixure of light and heavy ions,  then the light ions (e.g. protons) are quickly accelerated by this static field, and gain high energies on a distance of few micrometers.

This relativistic tweezer scheme has been demonstrated based on a three-dimensional particle-in-cell (PIC) simulation using code OSIRIS \cite{fonseca2002}, where two counter-propagating ten-cycle, temporally flat-top CP laser pulses with wavelengths ${\lambda_1 = 0.8}$~$\mu$m and ${\lambda_2 = 2\lambda_1}$, and ${a_{1,2} = a_0 = 16}$ are focused to 8~$\mu$m spots on a 20~nm hydrocarbon foil. The pre-ionized uniform plasma foil has electron density ${n_e = 100\, n_c}$, proton density  $10\ n_c$, and C$^{+6}$ ion density $15\ n_c$, where ${n_c=1.74\times10^{21}}$~cm$^{-3}$ is the critical plasma density for 800~nm laser wavelength. Initial electron temperature is set to 1 keV. To resolve kinetics of the high density plasma, we use numerical grid with the longitudinal cell size ${\Delta z = 3}$~nm and the transverse ones ${\Delta x =\Delta y= 10}$~nm. To ensure low particle noise, 32 macro-particles per species per cell are used. The output of this simulation is presented in Figure~\ref{fig1}.

Let us first consider the dynamics of an electron in the electromagnetic wave created by two lasers with the frequencies $\omega_{1}>\omega_{2}$, wavenumbers $k_{1\,,2}$, and the same amplitudes $a_0$. The field of interfering lasers has a phase component, ${a\propto \sin [(z- v_\text{tw}t) (k_1+k_2)/2]}$, which travels at a sub-luminal velocity ${v_\text{tw}= (\omega_1-\omega_2)/(k_1+k_2)}$, in the direction of the higher frequency laser. In a frame co-propagating with this beat-wave, both lasers have the same frequencies ${\omega_\text{tw} = \sqrt{ \omega_1 \omega_2 }}$ and wavenumbers ${k_\text{tw}=\sqrt{k_1 k_2}}$, and the total on-axis laser field can be written as,
\begin{equation}\label{tweezer_field}
 \mathbf{a}_\text{tw}=2 a_0\sin{k_\text{tw}z}(\mathbf{e}_y  \cos{\omega_\text{tw}}t + \mathbf{e}_x \sin{\omega_\text{tw}t})\,,
\end{equation}
where $\mathbf{e}_x$ and $\mathbf{e}_y$ are the transverse unit vectors. Considering that initially electron is at rest, its longitudinal motion follows the equation,
$${dp_z/dt = f_p - f_s} \,, $$
where two terms in the right-hand side are the ponderomotive force of the standing wave ${f_p=-(2a_0^2/\gamma_e) m_e c \omega_\text{tw} \sin 2 k_\text{tw} z}$, and the averaged Coulomb force from the charge-separation ${f_s=eE_z/2=2\pi e^2 n_el_0}$. Here, $l_0$ is the initial target thickness, and ${E_z=4\pi en_el_0}$ is the charge-separation field (shown with arrows in Fig.~\ref{fig1}b). To estimate the ratio between these forces, we first assume a minimal laser amplitude to be defined by the relativistic transparency condition \cite{vshivkov1998nonlinear}, $a_0\gtrsim 2\pi l_0 n_e/\lambda_1 n_c $, where we have added a fitting factor of 2 estimated from the simulations. We also assume the effective electron Lorentz factor $\gamma_e\approx a_0$, from which one can show that the charge-separation force is actually small compared to the maximum of ponderomotive force, $f_s\approx f_p/4$. Therefore, neglecting $f_s$, and for the initial electron velocity in the moving frame $v_{z0}=-v_\text{tw}$, we can estimate that electron is trapped, if ${v_\text{tw} \lesssim 2a_0 c/\sqrt{1+4a_0^2}}$. For the ultra-intense lasers, $a_0\gg 1$, this allows electron trapping even for the nearly luminal tweezer velocities, $v_\text{tw}\approx c$. In the present simulation, $v_\text{tw} =  c/3$.

\begin{figure*}[!ht]
  \centering
  \includegraphics[width=0.9\linewidth]{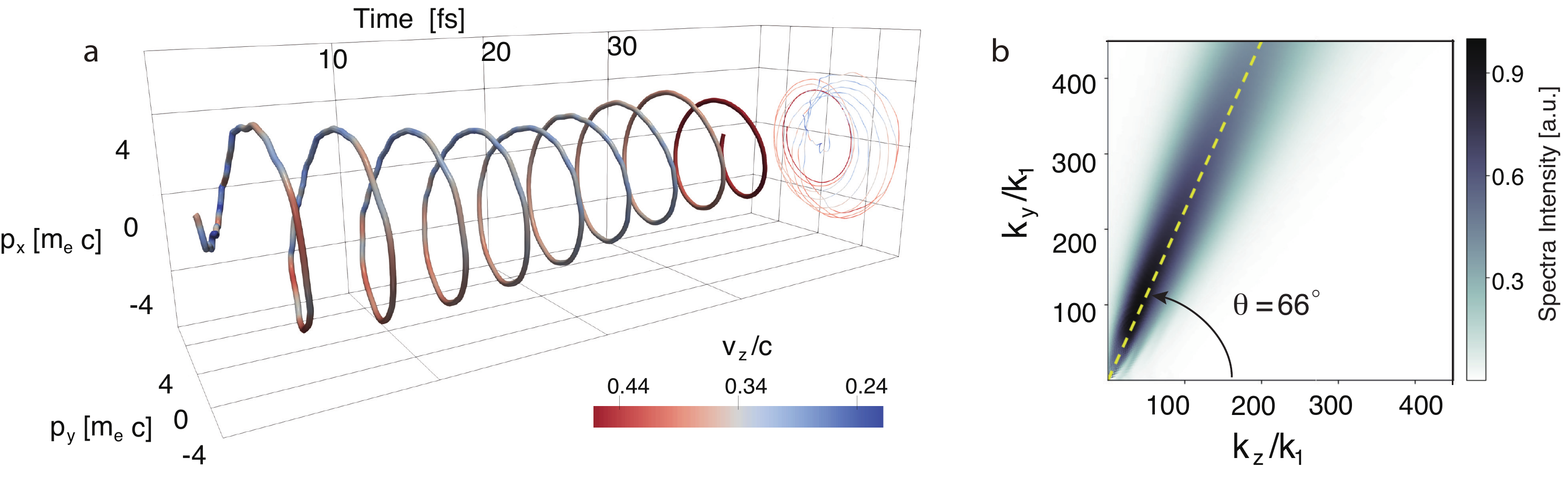}
  \caption{%\textbf{The trajectory of the tweezed electron disk centroid and characteristics of the emitted radiation.}
 (a), Time evolution of the average electron transverse momenta colored by its longitudinal velocity $v_z$. In the transverse plane, electrons move in a helical trajectory, and their longitudinal motion includes a constant velocity of 0.34c (group velocity of the beat pattern) and small oscillations.
 (b), The angular distribution of the incoherent radiated energy in normalized units. The dashed line shows the emission angle of $66^{\circ}$.}\label{fig2}
\end{figure*}

Once trapped, electron sheet moves helically in the laser field and travels with the beat-wave in a phase determined by both, laser ponderomotive force and static field of the ion plasma. Their averaged phase orbit is shown in Fig.~\ref{fig2}a. Since the transverse oscillations are relativistically strong, the electrons produce a broadband synchrotron light \cite{sokolov1968synchrotron}, which in this case is emitted into a ring-like shape at the angle ${\theta_\text{emit} = \arctan (\sqrt{c^2/v_\text{tw}^2-1})}$ to the tweezer propagation direction. In Fig.~\ref{fig2}b we show the numerically reconstructed spectral-angular radiation distribution, where the emission angle $\theta = 66^{\circ}$ is close to theoretical estimate $\theta_\text{emit}\simeq 70^{\circ}$. The critical photon energy $\hbar \omega_c \approx 0.25$ keV, and total radiated energy reaches a few micro-Joules level  (for the trapped charge $\simeq 20$ nC).

For the considered interaction parameters, the charge separation field between electron sheet and ion plasma reaches $E_z\sim$ 60 TV/m, and accelerates protons on a time scale, when the heavier carbon ions are essentially immobile (see Fig.\ref{fig1}b). This acceleration is nearly uniform, and it continues until protons outrun the electron sheet. The maximum acceleration time $t_\text{max}$ can be estimated as
\begin{eqnarray}
t_\text{max}=\frac{2 v_\text{tw} \gamma_{tw}^2}{\alpha}, \label{maximum_time_1D}
\end{eqnarray}
where $\alpha=eE_z/m_p$ is the acceleration rate of protons, $m_p$ is the proton rest mass, and $\gamma_\text{tw}=c/\sqrt{c^2-v_\text{tw}^2}$ is the Lorentz factor of the moving frame. Integrating the accelerated motion one can get the maximum energy gain of protons,
\begin{eqnarray}
\epsilon_\text{max} = 2 m_p v_{tw}^2 \gamma_{tw}^2. \label{maximum_energy_1D}
\end{eqnarray}

\begin{figure*}[!ht]
  \centering
 \includegraphics[width=\linewidth]{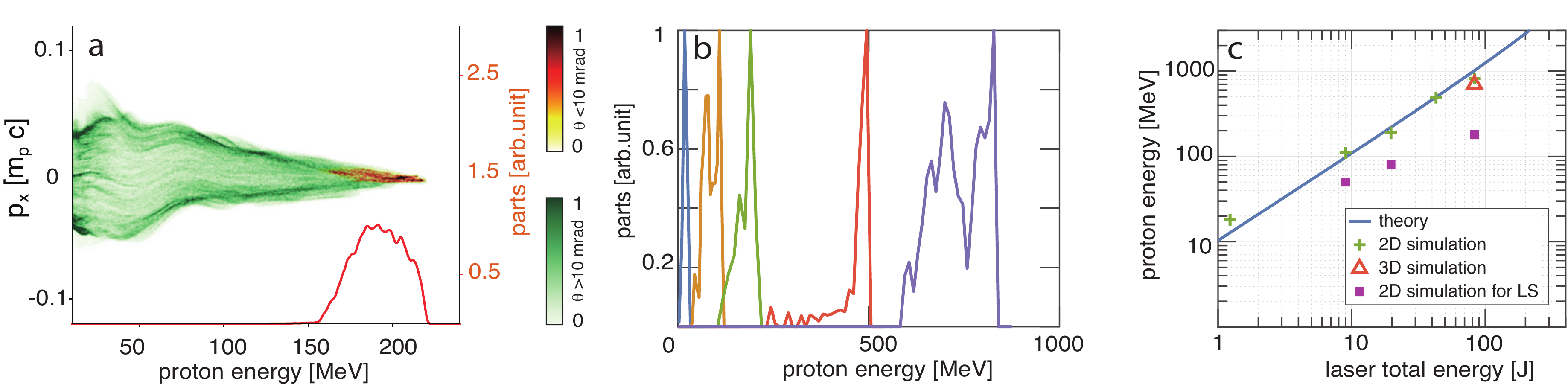}
  \caption{%\textbf{Proton phase space and its energy scaling.}
 (a), Density distribution of the accelerated protons in the ($P_x-\text{energy}$) space with divergence angles $\theta_{\perp}\equiv P_\perp/P_z$ less than 10 mrad (orange) and more than 10 mrad (green), where $m_p$ is proton rest mass. Energy spectrum of the collimated protons is plotted in orange curve.
 (b), Spectra of the collimated proton beams with $\theta_{\perp}\leq 10$ mrad at different laser energies of 1.2 J (blue), 9 J (orange), 19 J (green),  44 J (red) and 81 J (purple).
 (c), Scaling of the proton peak energy against the laser energy: solid blue curve shows the theoretical estimate; the green crosses and red triangles show 2D and 3D PIC simulation results, respectively; the purple squares gives 2D PIC simulation results for ``light sail'' (LS) scheme. }\label{fig3}
\end{figure*}

In the simulation presented in Fig.\ref{fig1}, acceleration process lasts for about 40~fs, and protons gain maximum energy of 220~MeV. These values are in good agreement with the theoretical estimates of 36~fs and 230~MeV from Eqs.~(\ref{maximum_time_1D}) and (\ref{maximum_energy_1D}) respectively. In Fig.~\ref{fig3}, we show the proton phase space at the moment when they outrun the electrons. One can see a quasi-monoenergetic group of protons with a FWHM energy spread around 20\% formed within a small divergence angle $\theta_{\perp}\equiv P_\perp/P_z\leq 10$ mrad (orange colormap and curve). This group contains $\sim10^{10}$ particles, and it may indeed be of interest for the radiotherapy applications \cite{bulanov2002CTpra,malka2004ct}. Meanwhile, about $10^{11}$ protons with the broadband spectrum and larger divergence, $\theta_{\perp}> 10$ mrad, (green in Fig.~\ref{fig3}a) originate from the borders of irradiated region, where the accelerating field becomes three dimensional (position dependent). 

We note, that the above analysis assumes the flat-top laser temporal profiles. It is important to verify how the discussed mechanisms are affected by the finite laser shapes, and for this we have run a series of additional 2D PIC simulations. These simulations show, that for beams with Gaussian temporal envelopes the tweezer effect is very similar to the idealized case, and production of 200 MeV protons in this case required slightly higher field amplitudes, $a_0=19$. The finite transverse profiles of laser beams also have an important effect on the scheme. Since the extracted electron sheet has a finite size, lasers can diffract at its edges, and this diffraction erodes the edges of the electron sheet, thus decreasing its size with propagation. From the simulations we have found that the time of the sheet destruction depends on the lasers radii as ${t_\text{eff} \approx 1.3 w_0/c}$.

From these considerations, we can now estimate how the maximum proton energy scales with the input laser energy. Erosion time of the electrons from the sheet edge defines the required laser durations as, ${\tau_{1\,,2}\simeq(1\mp v_\text{tw}/c)t_\text{eff}}$, and their energies scale as, ${W_{1\,,2} \propto \tau_{1,2} (a_0 w_0)^2 }$. Assuming that the optimal acceleration is reached, when $t_\text{eff}$ is close to the 1D maximum acceleration time $t_\text{max}$ (Eq.~\ref{maximum_time_1D}), one can find the scaling as:
\begin{eqnarray}
\epsilon_\text{max} [\text{MeV}]\simeq 0.34\left(\frac{W_\text{laser} \lambda_1}{w_0}\right)^2 + 49.5\frac{W_\text{laser} \lambda_1}{w_0}\, \,, \label{maximum_energy_3D}
\end{eqnarray}
where the total laser energy $W_\text{laser}=W_1+W_2$  is in Joules, and $\lambda_1$ and $w_0$ are in the same units. We note that the value of $\lambda_2$ is determined by the relation of $t_\text{eff}\simeq t_\text{max}$, and varies for different proton energy (see Eq.~\ref{maximum_energy_1D}).

We have tracked this dependency in a series of 2D PIC simulations, by considering only the collimated protons, in ${\theta_{\perp}\leq10}$~mrad (see Fig.~\ref{fig3}b). In these simulations, both lasers have Gaussian temporal and super-Gaussian transverse profiles, and target ion composition is the same as in the above 3D case. The values of $\lambda_1 = 0.8$~$\mu$m and $w_0=4\mu$m and $l_0=50$~nm are fixed, while other parameters were chosen in a way described previously in order to maintain the optimal acceleration conditions.

As shown in Fig.~\ref{fig3}b and \ref{fig3}c, protons with ${4\%-20\%}$ energy spreads and energies from 20 MeV to 830 MeV are obtained for 1 J to 80 J of laser energies. This agrees well with the dependency given by Eq.~(\ref{maximum_energy_3D}). Assuming a cylindrical symmetry we estimate the corresponding quantities of the physical protons in these simulations as $\sim 10^{10}$. In Fig.~\ref{fig3}c, the added 3D simulation also shows similar proton final energy. Besides, we have compared this tweezer approach with the ``light sail'' scheme for the same laser energy and optimal acceleration parameters \cite{yan2008,macchi2009}. The result shows, that the present scheme can provide 2-5 times higher energy gain than the ``light sail'' acceleration. By adding time jitters between laser pulses or imbalancing their amplitudes in the simulations, we have found that the acceptable jitter limit is about 10 femtoseconds, and the normalized laser amplitudes should not deviate from each other more than, $|a_2/a_1-1|\leq 0.2$.

In conclusion, we have proposed a high-intensity laser tweezer based on two relativistically intense lasers of different frequencies colliding on a nano-foil. The moving beat pattern of the lasers acts as a tweezer that drags the plasma electrons out of the ionized foil. The resultant capacitor-like field readily accelerates protons from the hydrocarbon plasma. The scheme is free of electron-ion coupling instability and is based on a simple nano-foil. This makes the proposed approach simpler to realize in practice compared to another recently proposed scheme that employs a standing wave incident on multi-nano-layers with an accurate phase control of a chirped laser \cite{MackenrothPRL2016}. The process is also accompanied by a strong synchrotron-like emission in the sub soft x-ray region, which can be used as a direct diagnostics of the interaction parameters, e.g. the tweezer phase velocity. Production of the two color laser pair, can be potentially achieved used through second-harmonic generation \cite{bargsten2017energy} or OPCPA techniques\cite{cerulloRSI2003, mikhailova2011ultra}. We anticipate this concept to open the path for the next generation of the compact laser-based high quality proton accelerators, and a wide range of their applications.

This work was supported by NSFC Grant No. 11425521, No. 11535006, No. 11475101 and No. 11775125, the Thousand Young Talents Program, and DOE grant DE-SC0010064. Simulations were performed on Sunway TaihuLight cluster at National Supercomputing Center and Edison cluster at NERSC. This wok has benefited from Gerry Schwartz and Heather Reisman, Israel Science Foundation, and VATAT supports.

\bibliography{tweezer_refs}

\begin{thebibliography}{34}%
\makeatletter
\providecommand \@ifxundefined [1]{%
 \@ifx{#1\undefined}
}%
\providecommand \@ifnum [1]{%
 \ifnum #1\expandafter \@firstoftwo
 \else \expandafter \@secondoftwo
 \fi
}%
\providecommand \@ifx [1]{%
 \ifx #1\expandafter \@firstoftwo
 \else \expandafter \@secondoftwo
 \fi
}%
\providecommand \natexlab [1]{#1}%
\providecommand \enquote  [1]{``#1''}%
\providecommand \bibnamefont  [1]{#1}%
\providecommand \bibfnamefont [1]{#1}%
\providecommand \citenamefont [1]{#1}%
\providecommand \href@noop [0]{\@secondoftwo}%
\providecommand \href [0]{\begingroup \@sanitize@url \@href}%
\providecommand \@href[1]{\@@startlink{#1}\@@href}%
\providecommand \@@href[1]{\endgroup#1\@@endlink}%
\providecommand \@sanitize@url [0]{\catcode `\\12\catcode `\$12\catcode
  `\&12\catcode `\#12\catcode `\^12\catcode `\_12\catcode `\%12\relax}%
\providecommand \@@startlink[1]{}%
\providecommand \@@endlink[0]{}%
\providecommand \url  [0]{\begingroup\@sanitize@url \@url }%
\providecommand \@url [1]{\endgroup\@href {#1}{\urlprefix }}%
\providecommand \urlprefix  [0]{URL }%
\providecommand \Eprint [0]{\href }%
\providecommand \doibase [0]{http://dx.doi.org/}%
\providecommand \selectlanguage [0]{\@gobble}%
\providecommand \bibinfo  [0]{\@secondoftwo}%
\providecommand \bibfield  [0]{\@secondoftwo}%
\providecommand \translation [1]{[#1]}%
\providecommand \BibitemOpen [0]{}%
\providecommand \bibitemStop [0]{}%
\providecommand \bibitemNoStop [0]{.\EOS\space}%
\providecommand \EOS [0]{\spacefactor3000\relax}%
\providecommand \BibitemShut  [1]{\csname bibitem#1\endcsname}%
\let\auto@bib@innerbib\@empty
%</preamble>
\bibitem [{\citenamefont {Ashkin}\ \emph {et~al.}(1986)\citenamefont {Ashkin},
  \citenamefont {Dziedzic}, \citenamefont {Bjorkholm},\ and\ \citenamefont
  {Chu}}]{ashkin1986}%
  \BibitemOpen
  \bibfield  {author} {\bibinfo {author} {\bibfnamefont {A.}~\bibnamefont
  {Ashkin}}, \bibinfo {author} {\bibfnamefont {J.~M.}\ \bibnamefont
  {Dziedzic}}, \bibinfo {author} {\bibfnamefont {J.~E.}\ \bibnamefont
  {Bjorkholm}}, \ and\ \bibinfo {author} {\bibfnamefont {S.}~\bibnamefont
  {Chu}},\ }\href@noop {} {\bibfield  {journal} {\bibinfo  {journal} {Optics
  Letters.}\ }\textbf {\bibinfo {volume} {11}},\ \bibinfo {pages} {288}
  (\bibinfo {year} {1986})}\BibitemShut {NoStop}%
\bibitem [{\citenamefont {Chu}(1991)}]{chu1991}%
  \BibitemOpen
  \bibfield  {author} {\bibinfo {author} {\bibfnamefont {S.}~\bibnamefont
  {Chu}},\ }\href@noop {} {\bibfield  {journal} {\bibinfo  {journal} {Science}\
  }\textbf {\bibinfo {volume} {253}},\ \bibinfo {pages} {861} (\bibinfo {year}
  {1991})}\BibitemShut {NoStop}%
\bibitem [{\citenamefont {Grier}(2003)}]{grier2003}%
  \BibitemOpen
  \bibfield  {author} {\bibinfo {author} {\bibfnamefont {D.~G.}\ \bibnamefont
  {Grier}},\ }\href@noop {} {\bibfield  {journal} {\bibinfo  {journal}
  {Nature}\ }\textbf {\bibinfo {volume} {424}},\ \bibinfo {pages} {810}
  (\bibinfo {year} {2003})}\BibitemShut {NoStop}%
\bibitem [{\citenamefont {Mourou}\ \emph {et~al.}(2006)\citenamefont {Mourou},
  \citenamefont {Tajima},\ and\ \citenamefont {Bulanov}}]{mourou2006optics}%
  \BibitemOpen
  \bibfield  {author} {\bibinfo {author} {\bibfnamefont {G.~A.}\ \bibnamefont
  {Mourou}}, \bibinfo {author} {\bibfnamefont {T.}~\bibnamefont {Tajima}}, \
  and\ \bibinfo {author} {\bibfnamefont {S.~V.}\ \bibnamefont {Bulanov}},\
  }\href@noop {} {\bibfield  {journal} {\bibinfo  {journal} {Reviews of Modern
  Physics}\ }\textbf {\bibinfo {volume} {78}},\ \bibinfo {pages} {309}
  (\bibinfo {year} {2006})}\BibitemShut {NoStop}%
\bibitem [{\citenamefont {Rosenbluth}\ and\ \citenamefont
  {Liu}(1972)}]{RosenbluthPRL1972}%
  \BibitemOpen
  \bibfield  {author} {\bibinfo {author} {\bibfnamefont {M.~N.}\ \bibnamefont
  {Rosenbluth}}\ and\ \bibinfo {author} {\bibfnamefont {C.~S.}\ \bibnamefont
  {Liu}},\ }\href@noop {} {\bibfield  {journal} {\bibinfo  {journal} {Physical
  Review Letters}\ }\textbf {\bibinfo {volume} {29}},\ \bibinfo {pages} {701}
  (\bibinfo {year} {1972})}\BibitemShut {NoStop}%
\bibitem [{\citenamefont {Tajima}\ and\ \citenamefont
  {Dawson}(1979)}]{tajima:PRL1979}%
  \BibitemOpen
  \bibfield  {author} {\bibinfo {author} {\bibfnamefont {T.}~\bibnamefont
  {Tajima}}\ and\ \bibinfo {author} {\bibfnamefont {J.~M.}\ \bibnamefont
  {Dawson}},\ }\href@noop {} {\bibfield  {journal} {\bibinfo  {journal}
  {Physical Review Letters}\ }\textbf {\bibinfo {volume} {43}},\ \bibinfo
  {pages} {267} (\bibinfo {year} {1979})}\BibitemShut {NoStop}%
\bibitem [{\citenamefont {Akhiezer}\ and\ \citenamefont
  {Polovin}(1956)}]{akhiezer1956theory}%
  \BibitemOpen
  \bibfield  {author} {\bibinfo {author} {\bibfnamefont {A.~I.}\ \bibnamefont
  {Akhiezer}}\ and\ \bibinfo {author} {\bibfnamefont {R.~V.}\ \bibnamefont
  {Polovin}},\ }\href@noop {} {\bibfield  {journal} {\bibinfo  {journal}
  {Soviet Phys. JETP}\ }\textbf {\bibinfo {volume} {3}},\ \bibinfo {pages}
  {696} (\bibinfo {year} {1956})}\BibitemShut {NoStop}%
\bibitem [{\citenamefont {Daido}\ \emph {et~al.}(2012)\citenamefont {Daido},
  \citenamefont {Nishiuchi},\ and\ \citenamefont
  {Pirozhkov}}]{daido2012review}%
  \BibitemOpen
  \bibfield  {author} {\bibinfo {author} {\bibfnamefont {H.}~\bibnamefont
  {Daido}}, \bibinfo {author} {\bibfnamefont {M.}~\bibnamefont {Nishiuchi}}, \
  and\ \bibinfo {author} {\bibfnamefont {A.~S.}\ \bibnamefont {Pirozhkov}},\
  }\href@noop {} {\bibfield  {journal} {\bibinfo  {journal} {Reports on
  Progress in Physics}\ }\textbf {\bibinfo {volume} {75}},\ \bibinfo {pages}
  {056401} (\bibinfo {year} {2012})}\BibitemShut {NoStop}%
\bibitem [{\citenamefont {Macchi}\ \emph {et~al.}(2013)\citenamefont {Macchi},
  \citenamefont {Borghesi},\ and\ \citenamefont {Passoni}}]{macchi2013ion}%
  \BibitemOpen
  \bibfield  {author} {\bibinfo {author} {\bibfnamefont {A.}~\bibnamefont
  {Macchi}}, \bibinfo {author} {\bibfnamefont {M.}~\bibnamefont {Borghesi}}, \
  and\ \bibinfo {author} {\bibfnamefont {M.}~\bibnamefont {Passoni}},\
  }\href@noop {} {\bibfield  {journal} {\bibinfo  {journal} {Reviews of Modern
  Physics}\ }\textbf {\bibinfo {volume} {85}},\ \bibinfo {pages} {751}
  (\bibinfo {year} {2013})}\BibitemShut {NoStop}%
\bibitem [{\citenamefont {Snavely}\ \emph {et~al.}(2000)\citenamefont
  {Snavely}, \citenamefont {Key}, \citenamefont {Hatchett}, \citenamefont
  {Cowan}, \citenamefont {Roth}, \citenamefont {Phillips}, \citenamefont
  {Stoyer}, \citenamefont {Henry}, \citenamefont {Sangster}, \citenamefont
  {Singh}, \citenamefont {Wilks}, \citenamefont {MacKinnon}, \citenamefont
  {Offenberger}, \citenamefont {Pennington}, \citenamefont {Yasuike},
  \citenamefont {Langdon}, \citenamefont {Lasinski}, \citenamefont {Johnson},
  \citenamefont {Perry},\ and\ \citenamefont {Campbell}}]{snavely2000TNSA}%
  \BibitemOpen
  \bibfield  {author} {\bibinfo {author} {\bibfnamefont {R.~A.}\ \bibnamefont
  {Snavely}}, \bibinfo {author} {\bibfnamefont {M.~H.}\ \bibnamefont {Key}},
  \bibinfo {author} {\bibfnamefont {S.~P.}\ \bibnamefont {Hatchett}}, \bibinfo
  {author} {\bibfnamefont {T.~E.}\ \bibnamefont {Cowan}}, \bibinfo {author}
  {\bibfnamefont {M.}~\bibnamefont {Roth}}, \bibinfo {author} {\bibfnamefont
  {T.~W.}\ \bibnamefont {Phillips}}, \bibinfo {author} {\bibfnamefont {M.~A.}\
  \bibnamefont {Stoyer}}, \bibinfo {author} {\bibfnamefont {E.~A.}\
  \bibnamefont {Henry}}, \bibinfo {author} {\bibfnamefont {T.~C.}\ \bibnamefont
  {Sangster}}, \bibinfo {author} {\bibfnamefont {M.~S.}\ \bibnamefont {Singh}},
  \bibinfo {author} {\bibfnamefont {S.~C.}\ \bibnamefont {Wilks}}, \bibinfo
  {author} {\bibfnamefont {A.}~\bibnamefont {MacKinnon}}, \bibinfo {author}
  {\bibfnamefont {A.}~\bibnamefont {Offenberger}}, \bibinfo {author}
  {\bibfnamefont {D.~M.}\ \bibnamefont {Pennington}}, \bibinfo {author}
  {\bibfnamefont {K.}~\bibnamefont {Yasuike}}, \bibinfo {author} {\bibfnamefont
  {A.~B.}\ \bibnamefont {Langdon}}, \bibinfo {author} {\bibfnamefont {B.~F.}\
  \bibnamefont {Lasinski}}, \bibinfo {author} {\bibfnamefont {J.}~\bibnamefont
  {Johnson}}, \bibinfo {author} {\bibfnamefont {M.~D.}\ \bibnamefont {Perry}},
  \ and\ \bibinfo {author} {\bibfnamefont {E.~M.}\ \bibnamefont {Campbell}},\
  }\href@noop {} {\bibfield  {journal} {\bibinfo  {journal} {Physical Review
  Letters}\ }\textbf {\bibinfo {volume} {85}},\ \bibinfo {pages} {2945}
  (\bibinfo {year} {2000})}\BibitemShut {NoStop}%
\bibitem [{\citenamefont {Esirkepov}\ \emph {et~al.}(2004)\citenamefont
  {Esirkepov}, \citenamefont {Borghesi}, \citenamefont {Bulanov}, \citenamefont
  {Mourou},\ and\ \citenamefont {Tajima}}]{esirkepov2004highly}%
  \BibitemOpen
  \bibfield  {author} {\bibinfo {author} {\bibfnamefont {T.}~\bibnamefont
  {Esirkepov}}, \bibinfo {author} {\bibfnamefont {M.}~\bibnamefont {Borghesi}},
  \bibinfo {author} {\bibfnamefont {S.}~\bibnamefont {Bulanov}}, \bibinfo
  {author} {\bibfnamefont {G.}~\bibnamefont {Mourou}}, \ and\ \bibinfo {author}
  {\bibfnamefont {T.}~\bibnamefont {Tajima}},\ }\href@noop {} {\bibfield
  {journal} {\bibinfo  {journal} {Physical Review Letters}\ }\textbf {\bibinfo
  {volume} {92}},\ \bibinfo {pages} {175003} (\bibinfo {year}
  {2004})}\BibitemShut {NoStop}%
\bibitem [{\citenamefont {Macchi}\ \emph {et~al.}(2005)\citenamefont {Macchi},
  \citenamefont {Cattani}, \citenamefont {Liseykina},\ and\ \citenamefont
  {Cornolti}}]{macchi2005}%
  \BibitemOpen
  \bibfield  {author} {\bibinfo {author} {\bibfnamefont {A.}~\bibnamefont
  {Macchi}}, \bibinfo {author} {\bibfnamefont {F.}~\bibnamefont {Cattani}},
  \bibinfo {author} {\bibfnamefont {T.~V.}\ \bibnamefont {Liseykina}}, \ and\
  \bibinfo {author} {\bibfnamefont {F.}~\bibnamefont {Cornolti}},\ }\href@noop
  {} {\bibfield  {journal} {\bibinfo  {journal} {Physical Review Letters}\
  }\textbf {\bibinfo {volume} {94}},\ \bibinfo {pages} {165003} (\bibinfo
  {year} {2005})}\BibitemShut {NoStop}%
\bibitem [{\citenamefont {Silva}\ \emph {et~al.}(2004)\citenamefont {Silva},
  \citenamefont {Marti}, \citenamefont {Davies}, \citenamefont {Fonseca},
  \citenamefont {Ren}, \citenamefont {Tsung},\ and\ \citenamefont
  {Mori}}]{silva2004}%
  \BibitemOpen
  \bibfield  {author} {\bibinfo {author} {\bibfnamefont {L.~O.}\ \bibnamefont
  {Silva}}, \bibinfo {author} {\bibfnamefont {M.}~\bibnamefont {Marti}},
  \bibinfo {author} {\bibfnamefont {J.~R.}\ \bibnamefont {Davies}}, \bibinfo
  {author} {\bibfnamefont {R.~A.}\ \bibnamefont {Fonseca}}, \bibinfo {author}
  {\bibfnamefont {C.}~\bibnamefont {Ren}}, \bibinfo {author} {\bibfnamefont
  {F.~S.}\ \bibnamefont {Tsung}}, \ and\ \bibinfo {author} {\bibfnamefont
  {W.~B.}\ \bibnamefont {Mori}},\ }\href@noop {} {\bibfield  {journal}
  {\bibinfo  {journal} {Physical Review Letters}\ }\textbf {\bibinfo {volume}
  {92}},\ \bibinfo {pages} {015002} (\bibinfo {year} {2004})}\BibitemShut
  {NoStop}%
\bibitem [{\citenamefont {Haberberger}\ \emph {et~al.}(2012)\citenamefont
  {Haberberger}, \citenamefont {Tochitsky}, \citenamefont {Fiuza},
  \citenamefont {Gong}, \citenamefont {Fonseca}, \citenamefont {Silva},
  \citenamefont {Mori},\ and\ \citenamefont {Joshi}}]{haberberger2012}%
  \BibitemOpen
  \bibfield  {author} {\bibinfo {author} {\bibfnamefont {D.}~\bibnamefont
  {Haberberger}}, \bibinfo {author} {\bibfnamefont {S.}~\bibnamefont
  {Tochitsky}}, \bibinfo {author} {\bibfnamefont {F.}~\bibnamefont {Fiuza}},
  \bibinfo {author} {\bibfnamefont {C.}~\bibnamefont {Gong}}, \bibinfo {author}
  {\bibfnamefont {R.~A.}\ \bibnamefont {Fonseca}}, \bibinfo {author}
  {\bibfnamefont {L.~O.}\ \bibnamefont {Silva}}, \bibinfo {author}
  {\bibfnamefont {W.~B.}\ \bibnamefont {Mori}}, \ and\ \bibinfo {author}
  {\bibfnamefont {C.}~\bibnamefont {Joshi}},\ }\href@noop {} {\bibfield
  {journal} {\bibinfo  {journal} {Nature Physics}\ }\textbf {\bibinfo {volume}
  {8}},\ \bibinfo {pages} {95} (\bibinfo {year} {2012})}\BibitemShut {NoStop}%
\bibitem [{\citenamefont {Fiuza}\ \emph {et~al.}(2012)\citenamefont {Fiuza},
  \citenamefont {Stockem}, \citenamefont {Boella}, \citenamefont {Fonseca},
  \citenamefont {Silva}, \citenamefont {Haberberger}, \citenamefont
  {Tochitsky}, \citenamefont {Gong}, \citenamefont {Mori},\ and\ \citenamefont
  {Joshi}}]{fiuza2012prl}%
  \BibitemOpen
  \bibfield  {author} {\bibinfo {author} {\bibfnamefont {F.}~\bibnamefont
  {Fiuza}}, \bibinfo {author} {\bibfnamefont {A.}~\bibnamefont {Stockem}},
  \bibinfo {author} {\bibfnamefont {E.}~\bibnamefont {Boella}}, \bibinfo
  {author} {\bibfnamefont {R.~A.}\ \bibnamefont {Fonseca}}, \bibinfo {author}
  {\bibfnamefont {L.~O.}\ \bibnamefont {Silva}}, \bibinfo {author}
  {\bibfnamefont {D.}~\bibnamefont {Haberberger}}, \bibinfo {author}
  {\bibfnamefont {S.}~\bibnamefont {Tochitsky}}, \bibinfo {author}
  {\bibfnamefont {C.}~\bibnamefont {Gong}}, \bibinfo {author} {\bibfnamefont
  {W.~B.}\ \bibnamefont {Mori}}, \ and\ \bibinfo {author} {\bibfnamefont
  {C.}~\bibnamefont {Joshi}},\ }\href@noop {} {\bibfield  {journal} {\bibinfo
  {journal} {Physical Review Letters}\ }\textbf {\bibinfo {volume} {109}},\
  \bibinfo {pages} {215001} (\bibinfo {year} {2012})}\BibitemShut {NoStop}%
\bibitem [{\citenamefont {Bulanov}\ \emph {et~al.}(2005)\citenamefont
  {Bulanov}, \citenamefont {Dylov}, \citenamefont {Esirkepov}, \citenamefont
  {Kamenets},\ and\ \citenamefont {Sokolov}}]{bulanov2005}%
  \BibitemOpen
  \bibfield  {author} {\bibinfo {author} {\bibfnamefont {S.~V.}\ \bibnamefont
  {Bulanov}}, \bibinfo {author} {\bibfnamefont {D.~V.}\ \bibnamefont {Dylov}},
  \bibinfo {author} {\bibfnamefont {T.~Z.}\ \bibnamefont {Esirkepov}}, \bibinfo
  {author} {\bibfnamefont {F.~F.}\ \bibnamefont {Kamenets}}, \ and\ \bibinfo
  {author} {\bibfnamefont {D.~V.}\ \bibnamefont {Sokolov}},\ }\href@noop {}
  {\bibfield  {journal} {\bibinfo  {journal} {Plasma Physics Reports}\ }\textbf
  {\bibinfo {volume} {31}},\ \bibinfo {pages} {369} (\bibinfo {year}
  {2005})}\BibitemShut {NoStop}%
\bibitem [{\citenamefont {Bulanov}\ \emph {et~al.}(2010)\citenamefont
  {Bulanov}, \citenamefont {Bychenkov}, \citenamefont {Chvykov}, \citenamefont
  {Kalinchenko}, \citenamefont {Litzenberg}, \citenamefont {Matsuoka},
  \citenamefont {Thomas}, \citenamefont {Willingale}, \citenamefont {Yanovsky},
  \citenamefont {Krushelnick},\ and\ \citenamefont
  {Maksimchuk}}]{bulanov2010pop}%
  \BibitemOpen
  \bibfield  {author} {\bibinfo {author} {\bibfnamefont {S.~S.}\ \bibnamefont
  {Bulanov}}, \bibinfo {author} {\bibfnamefont {V.~Y.}\ \bibnamefont
  {Bychenkov}}, \bibinfo {author} {\bibfnamefont {V.}~\bibnamefont {Chvykov}},
  \bibinfo {author} {\bibfnamefont {G.}~\bibnamefont {Kalinchenko}}, \bibinfo
  {author} {\bibfnamefont {D.~W.}\ \bibnamefont {Litzenberg}}, \bibinfo
  {author} {\bibfnamefont {T.}~\bibnamefont {Matsuoka}}, \bibinfo {author}
  {\bibfnamefont {A.~G.~R.}\ \bibnamefont {Thomas}}, \bibinfo {author}
  {\bibfnamefont {L.}~\bibnamefont {Willingale}}, \bibinfo {author}
  {\bibfnamefont {V.}~\bibnamefont {Yanovsky}}, \bibinfo {author}
  {\bibfnamefont {K.}~\bibnamefont {Krushelnick}}, \ and\ \bibinfo {author}
  {\bibfnamefont {A.}~\bibnamefont {Maksimchuk}},\ }\href@noop {} {\bibfield
  {journal} {\bibinfo  {journal} {Physics of Plasmas}\ }\textbf {\bibinfo
  {volume} {17}},\ \bibinfo {pages} {043105} (\bibinfo {year}
  {2010})}\BibitemShut {NoStop}%
\bibitem [{\citenamefont {Nakamura}\ \emph {et~al.}(2010)\citenamefont
  {Nakamura}, \citenamefont {Bulanov}, \citenamefont {Esirkepov},\ and\
  \citenamefont {Kando}}]{nakamura2010}%
  \BibitemOpen
  \bibfield  {author} {\bibinfo {author} {\bibfnamefont {T.}~\bibnamefont
  {Nakamura}}, \bibinfo {author} {\bibfnamefont {S.~V.}\ \bibnamefont
  {Bulanov}}, \bibinfo {author} {\bibfnamefont {T.~Z.}\ \bibnamefont
  {Esirkepov}}, \ and\ \bibinfo {author} {\bibfnamefont {M.}~\bibnamefont
  {Kando}},\ }\href@noop {} {\bibfield  {journal} {\bibinfo  {journal}
  {Physical Review Letters}\ }\textbf {\bibinfo {volume} {105}},\ \bibinfo
  {pages} {135002} (\bibinfo {year} {2010})}\BibitemShut {NoStop}%
\bibitem [{\citenamefont {Klimo}\ \emph {et~al.}(2008)\citenamefont {Klimo},
  \citenamefont {Psikal}, \citenamefont {Limpouch},\ and\ \citenamefont
  {Tikhonchuk}}]{klimo2008}%
  \BibitemOpen
  \bibfield  {author} {\bibinfo {author} {\bibfnamefont {O.}~\bibnamefont
  {Klimo}}, \bibinfo {author} {\bibfnamefont {J.}~\bibnamefont {Psikal}},
  \bibinfo {author} {\bibfnamefont {J.}~\bibnamefont {Limpouch}}, \ and\
  \bibinfo {author} {\bibfnamefont {V.~T.}\ \bibnamefont {Tikhonchuk}},\
  }\href@noop {} {\bibfield  {journal} {\bibinfo  {journal} {Phys. Rev. ST
  Accel. Beams}\ }\textbf {\bibinfo {volume} {11}},\ \bibinfo {pages} {031301}
  (\bibinfo {year} {2008})}\BibitemShut {NoStop}%
\bibitem [{\citenamefont {Robinson}\ \emph {et~al.}(2008)\citenamefont
  {Robinson}, \citenamefont {Zepf}, \citenamefont {Kar}, \citenamefont
  {Evans},\ and\ \citenamefont {Bellei}}]{robinson2008}%
  \BibitemOpen
  \bibfield  {author} {\bibinfo {author} {\bibfnamefont {A.~P.~L.}\
  \bibnamefont {Robinson}}, \bibinfo {author} {\bibfnamefont {M.}~\bibnamefont
  {Zepf}}, \bibinfo {author} {\bibfnamefont {S.}~\bibnamefont {Kar}}, \bibinfo
  {author} {\bibfnamefont {R.~G.}\ \bibnamefont {Evans}}, \ and\ \bibinfo
  {author} {\bibfnamefont {C.}~\bibnamefont {Bellei}},\ }\href@noop {}
  {\bibfield  {journal} {\bibinfo  {journal} {New J. Phys.}\ }\textbf {\bibinfo
  {volume} {10}},\ \bibinfo {pages} {013021} (\bibinfo {year}
  {2008})}\BibitemShut {NoStop}%
\bibitem [{\citenamefont {Yan}\ \emph {et~al.}(2008)\citenamefont {Yan} \emph
  {et~al.}}]{yan2008}%
  \BibitemOpen
  \bibfield  {author} {\bibinfo {author} {\bibfnamefont {X.~Q.}\ \bibnamefont
  {Yan}} \emph {et~al.},\ }\href@noop {} {\bibfield  {journal} {\bibinfo
  {journal} {Physical Review Letters}\ }\textbf {\bibinfo {volume} {100}},\
  \bibinfo {pages} {135003} (\bibinfo {year} {2008})}\BibitemShut {NoStop}%
\bibitem [{\citenamefont {Macchi}\ \emph {et~al.}(2009)\citenamefont {Macchi},
  \citenamefont {Veghini},\ and\ \citenamefont {Pegoraro}}]{macchi2009}%
  \BibitemOpen
  \bibfield  {author} {\bibinfo {author} {\bibfnamefont {A.}~\bibnamefont
  {Macchi}}, \bibinfo {author} {\bibfnamefont {S.}~\bibnamefont {Veghini}}, \
  and\ \bibinfo {author} {\bibfnamefont {F.}~\bibnamefont {Pegoraro}},\
  }\href@noop {} {\bibfield  {journal} {\bibinfo  {journal} {Physical Review
  Letters}\ }\textbf {\bibinfo {volume} {103}},\ \bibinfo {pages} {085003}
  (\bibinfo {year} {2009})}\BibitemShut {NoStop}%
\bibitem [{\citenamefont {Pegoraro}\ and\ \citenamefont
  {Bulanov}(2007)}]{pegoraro2007}%
  \BibitemOpen
  \bibfield  {author} {\bibinfo {author} {\bibfnamefont {F.}~\bibnamefont
  {Pegoraro}}\ and\ \bibinfo {author} {\bibfnamefont {S.~V.}\ \bibnamefont
  {Bulanov}},\ }\href@noop {} {\bibfield  {journal} {\bibinfo  {journal}
  {Physical Review Letters}\ }\textbf {\bibinfo {volume} {99}},\ \bibinfo
  {pages} {065002} (\bibinfo {year} {2007})}\BibitemShut {NoStop}%
\bibitem [{\citenamefont {Wan}\ \emph {et~al.}(2016)\citenamefont {Wan} \emph
  {et~al.}}]{wan2016}%
  \BibitemOpen
  \bibfield  {author} {\bibinfo {author} {\bibfnamefont {Y.}~\bibnamefont
  {Wan}} \emph {et~al.},\ }\href@noop {} {\bibfield  {journal} {\bibinfo
  {journal} {Physical Review Letters}\ }\textbf {\bibinfo {volume} {117}},\
  \bibinfo {pages} {234801} (\bibinfo {year} {2016})}\BibitemShut {NoStop}%
\bibitem [{\citenamefont {Wan}\ \emph {et~al.}(2018)\citenamefont {Wan},
  \citenamefont {Pai}, \citenamefont {Zhang}, \citenamefont {Li}, \citenamefont
  {Wu}, \citenamefont {Hua}, \citenamefont {Lu}, \citenamefont {Joshi},
  \citenamefont {Mori},\ and\ \citenamefont {Malka}}]{wan2018physical}%
  \BibitemOpen
  \bibfield  {author} {\bibinfo {author} {\bibfnamefont {Y.}~\bibnamefont
  {Wan}}, \bibinfo {author} {\bibfnamefont {C.-H.}\ \bibnamefont {Pai}},
  \bibinfo {author} {\bibfnamefont {C.}~\bibnamefont {Zhang}}, \bibinfo
  {author} {\bibfnamefont {F.}~\bibnamefont {Li}}, \bibinfo {author}
  {\bibfnamefont {Y.}~\bibnamefont {Wu}}, \bibinfo {author} {\bibfnamefont
  {J.}~\bibnamefont {Hua}}, \bibinfo {author} {\bibfnamefont {W.}~\bibnamefont
  {Lu}}, \bibinfo {author} {\bibfnamefont {C.}~\bibnamefont {Joshi}}, \bibinfo
  {author} {\bibfnamefont {W.}~\bibnamefont {Mori}}, \ and\ \bibinfo {author}
  {\bibfnamefont {V.}~\bibnamefont {Malka}},\ }\href@noop {} {\bibfield
  {journal} {\bibinfo  {journal} {Physical Review E}\ }\textbf {\bibinfo
  {volume} {98}},\ \bibinfo {pages} {013202} (\bibinfo {year}
  {2018})}\BibitemShut {NoStop}%
\bibitem [{\citenamefont {Fonseca}\ \emph {et~al.}(2002)\citenamefont {Fonseca}
  \emph {et~al.}}]{fonseca2002}%
  \BibitemOpen
  \bibfield  {author} {\bibinfo {author} {\bibfnamefont {R.~A.}\ \bibnamefont
  {Fonseca}} \emph {et~al.}\ }(\bibinfo  {publisher} {Springer-Verlag Berlin},\
  \bibinfo {year} {2002})\ pp.\ \bibinfo {pages} {342--351}\BibitemShut
  {NoStop}%
\bibitem [{\citenamefont {Vshivkov}\ \emph {et~al.}(1998)\citenamefont
  {Vshivkov}, \citenamefont {Naumova}, \citenamefont {Pegoraro},\ and\
  \citenamefont {Bulanov}}]{vshivkov1998nonlinear}%
  \BibitemOpen
  \bibfield  {author} {\bibinfo {author} {\bibfnamefont {V.~A.}\ \bibnamefont
  {Vshivkov}}, \bibinfo {author} {\bibfnamefont {N.~M.}\ \bibnamefont
  {Naumova}}, \bibinfo {author} {\bibfnamefont {F.}~\bibnamefont {Pegoraro}}, \
  and\ \bibinfo {author} {\bibfnamefont {S.}~\bibnamefont {Bulanov}},\
  }\href@noop {} {\bibfield  {journal} {\bibinfo  {journal} {Physics of
  Plasmas}\ }\textbf {\bibinfo {volume} {5}},\ \bibinfo {pages} {2727}
  (\bibinfo {year} {1998})}\BibitemShut {NoStop}%
\bibitem [{\citenamefont {Sokolov}\ and\ \citenamefont
  {Ternov}(1968)}]{sokolov1968synchrotron}%
  \BibitemOpen
  \bibfield  {author} {\bibinfo {author} {\bibfnamefont {A.}~\bibnamefont
  {Sokolov}}\ and\ \bibinfo {author} {\bibfnamefont {I.}~\bibnamefont
  {Ternov}},\ }\href@noop {} {\emph {\bibinfo {title} {Synchrotron
  Radiation}}}\ (\bibinfo  {publisher} {Pergamon Press},\ \bibinfo {year}
  {1968})\BibitemShut {NoStop}%
\bibitem [{\citenamefont {Bulanov}\ \emph {et~al.}(2002)\citenamefont
  {Bulanov}, \citenamefont {Esirkepov}, \citenamefont {Khoroshkov},
  \citenamefont {Kunetsov},\ and\ \citenamefont {Pegoraro}}]{bulanov2002CTpra}%
  \BibitemOpen
  \bibfield  {author} {\bibinfo {author} {\bibfnamefont {S.~V.}\ \bibnamefont
  {Bulanov}}, \bibinfo {author} {\bibfnamefont {T.~Z.}\ \bibnamefont
  {Esirkepov}}, \bibinfo {author} {\bibfnamefont {V.~S.}\ \bibnamefont
  {Khoroshkov}}, \bibinfo {author} {\bibfnamefont {A.~V.}\ \bibnamefont
  {Kunetsov}}, \ and\ \bibinfo {author} {\bibfnamefont {F.}~\bibnamefont
  {Pegoraro}},\ }\href@noop {} {\bibfield  {journal} {\bibinfo  {journal}
  {Physical Review A}\ }\textbf {\bibinfo {volume} {299}},\ \bibinfo {pages}
  {240} (\bibinfo {year} {2002})}\BibitemShut {NoStop}%
\bibitem [{\citenamefont {Malka}\ \emph {et~al.}(2004)\citenamefont {Malka}
  \emph {et~al.}}]{malka2004ct}%
  \BibitemOpen
  \bibfield  {author} {\bibinfo {author} {\bibfnamefont {V.}~\bibnamefont
  {Malka}} \emph {et~al.},\ }\href@noop {} {\bibfield  {journal} {\bibinfo
  {journal} {Medical Physics}\ }\textbf {\bibinfo {volume} {31}},\ \bibinfo
  {pages} {1587} (\bibinfo {year} {2004})}\BibitemShut {NoStop}%
\bibitem [{\citenamefont {Mackenroth}\ \emph {et~al.}(2016)\citenamefont
  {Mackenroth}, \citenamefont {Gonoskov},\ and\ \citenamefont
  {Marklund}}]{MackenrothPRL2016}%
  \BibitemOpen
  \bibfield  {author} {\bibinfo {author} {\bibfnamefont {F.}~\bibnamefont
  {Mackenroth}}, \bibinfo {author} {\bibfnamefont {A.}~\bibnamefont
  {Gonoskov}}, \ and\ \bibinfo {author} {\bibfnamefont {M.}~\bibnamefont
  {Marklund}},\ }\href@noop {} {\bibfield  {journal} {\bibinfo  {journal}
  {Physical Review Letters}\ }\textbf {\bibinfo {volume} {117}},\ \bibinfo
  {pages} {104801} (\bibinfo {year} {2016})}\BibitemShut {NoStop}%
\bibitem [{\citenamefont {Bargsten}\ \emph {et~al.}(2017)\citenamefont
  {Bargsten}, \citenamefont {Hollinger}, \citenamefont {Capeluto},
  \citenamefont {Kaymak}, \citenamefont {Pukhov}, \citenamefont {Wang},
  \citenamefont {Rockwood}, \citenamefont {Wang}, \citenamefont {Keiss},
  \citenamefont {Tommasini} \emph {et~al.}}]{bargsten2017energy}%
  \BibitemOpen
  \bibfield  {author} {\bibinfo {author} {\bibfnamefont {C.}~\bibnamefont
  {Bargsten}}, \bibinfo {author} {\bibfnamefont {R.}~\bibnamefont {Hollinger}},
  \bibinfo {author} {\bibfnamefont {M.~G.}\ \bibnamefont {Capeluto}}, \bibinfo
  {author} {\bibfnamefont {V.}~\bibnamefont {Kaymak}}, \bibinfo {author}
  {\bibfnamefont {A.}~\bibnamefont {Pukhov}}, \bibinfo {author} {\bibfnamefont
  {S.}~\bibnamefont {Wang}}, \bibinfo {author} {\bibfnamefont {A.}~\bibnamefont
  {Rockwood}}, \bibinfo {author} {\bibfnamefont {Y.}~\bibnamefont {Wang}},
  \bibinfo {author} {\bibfnamefont {D.}~\bibnamefont {Keiss}}, \bibinfo
  {author} {\bibfnamefont {R.}~\bibnamefont {Tommasini}},  \emph {et~al.},\
  }\href@noop {} {\bibfield  {journal} {\bibinfo  {journal} {Science Advances}\
  }\textbf {\bibinfo {volume} {3}},\ \bibinfo {pages} {e1601558} (\bibinfo
  {year} {2017})}\BibitemShut {NoStop}%
\bibitem [{\citenamefont {Cerullo}\ and\ \citenamefont
  {De~Silvestri}(2003)}]{cerulloRSI2003}%
  \BibitemOpen
  \bibfield  {author} {\bibinfo {author} {\bibfnamefont {G.}~\bibnamefont
  {Cerullo}}\ and\ \bibinfo {author} {\bibfnamefont {S.}~\bibnamefont
  {De~Silvestri}},\ }\href@noop {} {\bibfield  {journal} {\bibinfo  {journal}
  {Review of Scientific Instruments}\ }\textbf {\bibinfo {volume} {74}},\
  \bibinfo {pages} {1} (\bibinfo {year} {2003})}\BibitemShut {NoStop}%
\bibitem [{\citenamefont {Mikhailova}\ \emph {et~al.}(2011)\citenamefont
  {Mikhailova}, \citenamefont {Buck}, \citenamefont {Borot}, \citenamefont
  {Schmid}, \citenamefont {Sears}, \citenamefont {Tsakiris}, \citenamefont
  {Krausz},\ and\ \citenamefont {Veisz}}]{mikhailova2011ultra}%
  \BibitemOpen
  \bibfield  {author} {\bibinfo {author} {\bibfnamefont {J.~M.}\ \bibnamefont
  {Mikhailova}}, \bibinfo {author} {\bibfnamefont {A.}~\bibnamefont {Buck}},
  \bibinfo {author} {\bibfnamefont {A.}~\bibnamefont {Borot}}, \bibinfo
  {author} {\bibfnamefont {K.}~\bibnamefont {Schmid}}, \bibinfo {author}
  {\bibfnamefont {C.}~\bibnamefont {Sears}}, \bibinfo {author} {\bibfnamefont
  {G.~D.}\ \bibnamefont {Tsakiris}}, \bibinfo {author} {\bibfnamefont
  {F.}~\bibnamefont {Krausz}}, \ and\ \bibinfo {author} {\bibfnamefont
  {L.}~\bibnamefont {Veisz}},\ }\href@noop {} {\bibfield  {journal} {\bibinfo
  {journal} {Optics Letters}\ }\textbf {\bibinfo {volume} {36}},\ \bibinfo
  {pages} {3145} (\bibinfo {year} {2011})}\BibitemShut {NoStop}%
\end{thebibliography}%
\end{document}